\documentclass{kluwer}    
\usepackage{graphicx}

\begin{document}
\begin{article}
\begin{opening}         
\title{Galaxy Threshing and Ultra-Compact Dwarfs in the Fornax Cluster}
\author{Michael D. \surname{Gregg},
Michael J. \surname{Drinkwater}, Michael J. \surname{Hilker},\\
Steven \surname{Phillipps},
J. Bryn \surname{Jones}, \& Henry C. \surname{Ferguson}
}
\runningauthor{Michael D. Gregg, et al.}
\runningtitle{Ultra-Compact Dwarfs}
\institute{Univ.\ of California, Davis, and Inst. for Geophysics and
            Planetary Physics, Lawrence Livermore National Laboratory;
Univ.\ of Queensland;
Univ.\ Bonn; 
Univ.\ of Bristol; 
Univ.\ of Nottingham; \& 
STScI}

\begin{abstract}
We have discovered a new type of galaxy in the Fornax
Cluster: ``ultra-compact'' dwarfs (UCDs).  The UCDs are
unresolved in ground-based imaging and have spectra typical of old
stellar systems.  Although the UCDs resemble overgrown globular
clusters, based on VLT UVES echelle spectroscopy, they appear to be
dynamically distinct systems with higher internal velocity dispersions
and M/L ratios for a given luminosity than Milky Way or M31 globulars.
Our preferred explanation for their origin is that they are the remnant
nuclei of dwarf elliptical galaxies which have been tidally stripped,
or ``threshed'' by repeated encounters with the central cluster
galaxy, NGC1399.  If correct, then tidal stripping of nucleated dwarfs
to form UCDs may, over a Hubble time, be an important source of the
plentiful globular cluster population in the halo of NGC1399, and, by
implication, other cD galaxies.  In this picture, the dwarf elliptical
halo contents, up to 99\% of the original dwarf luminosity,
contribute a significant fraction of the populations of intergalactic
stars, globulars, and gas in galaxy clusters.
\end{abstract}

\end{opening}           

\vspace {-0.25in}
\section{Introduction}  
\vspace {-0.15in}

Dwarf galaxies account for only a small fraction of the total mass and
light of galaxy clusters; nevertheless, they are a key element in the
evolution of galaxy clusters.  Evidence is gathering that evolution of
rich clusters can be gauged by its present population of dwarf
galaxies.  Tidal interactions which whittle away at (Moore et al.\
1996, Bekki, Couch, \& Drinkwater 2001a) and even disrupt entire
systems (Gregg \& West 1998) can liberate debris from bright galaxies,
giving rise to large populations of dwarf galaxies (Lopez-Cruz et al.\
1996).  In the opposite direction, dwarfs may lose their individual
identities by accreting onto the central elliptical, creating cD
galaxies.  This ebb and flow of the dwarf galaxy population may be
episodic and a measure of the dynamical state of a cluster (Lopez-Cruz
et al.\ 1996), and a complete understanding of galaxy clusters must
incorporate not only the present state of dwarf galaxies but also
their life histories.

The {\em Fornax Cluster Spectroscopic Survey} (FCSS; Drinkwater et
al.\ 2000a) is using the 2-Degree Field 400-fiber spectrograph on the
Anglo-Australian Telescope to obtain spectra for every object down to
$B_J<19.7$ in a $4^\circ \times 3^\circ$ area centered on the Fornax
cluster, about 14000 targets in all.  The FCSS is unique in targeting
{\em all objects, stellar and nonstellar} on Schmidt plates.  This
complete sample of galaxies over a large range of magnitude and
surface brightness will permit study of the luminosity function and
dynamics of Fornax in greater detail than has been done previously
(e.g., Ferguson 1989).  Now $\sim 70\%$ finished, the most exciting
find from the FCSS is the identification of seven objects from the
unresolved ``stellar'' targets which are ``ultra-compact dwarf'' (UCD)
cluster members with $-13 < M_B < -11$, and sizes $\lesssim 100$~pc.

\begin{figure}[t!]
\includegraphics[width=5in]{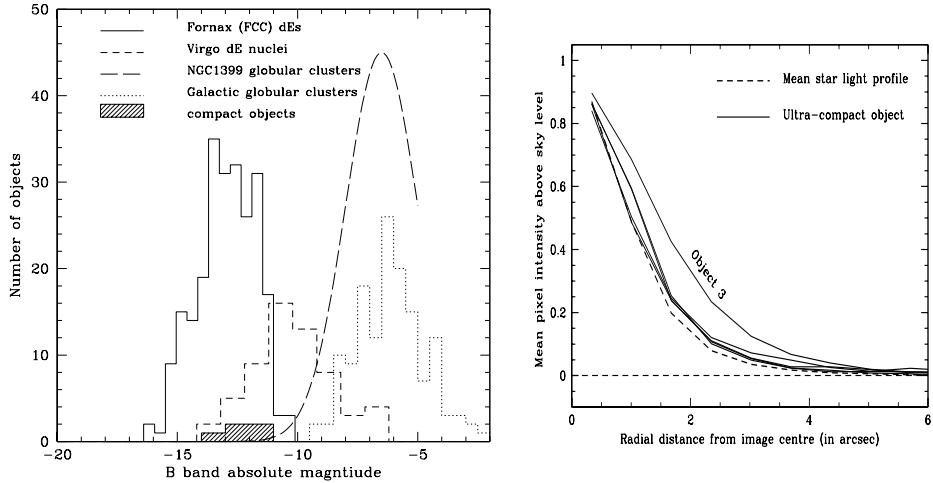}
\caption{\small {\bf Left:} 
The UCDs placed in luminosity context with more familiar
objects.  {\bf Right:}
Light profiles from AAT CCD images for
5 UCDs compared to a star in 1.2'' seeing.  All but one
is indistinguishable from a point source; consequently, such objects
are never discovered in conventional galaxy redshift surveys.}
\end{figure}

\vspace {-0.25in}
\section{UCD Properties}
\vspace {-0.15in}

Distributed widely across the cluster, the UCDs are not, as a class,
associated with bright galaxies (Drinkwater et al.\ 2000b; Phillipps
et al.\ 2001).  The UCDs
are $\sim 2$ magnitudes brighter than the largest globular clusters in
NGC~1399  (Figure~1) and have absorption-line spectra indicative of older,
moderately metal-poor stellar populations.  All are classified as
stellar in Schmidt plate material, which provides the target lists for
the FCSS, and all but the brightest UCD are still essentially
unresolved in ground-based CCD images in $\sim 1.2''$ seeing (Figure~1).
Such objects are passed over by galaxy redshift surveys,
rejected because of their stellar appearance.  Because the 7 UCDs are
near our spectroscopic survey magnitude limit, we predict that many
fainter UCDs exist in the cluster.

\begin{figure}[t!]
\includegraphics[width=5in]{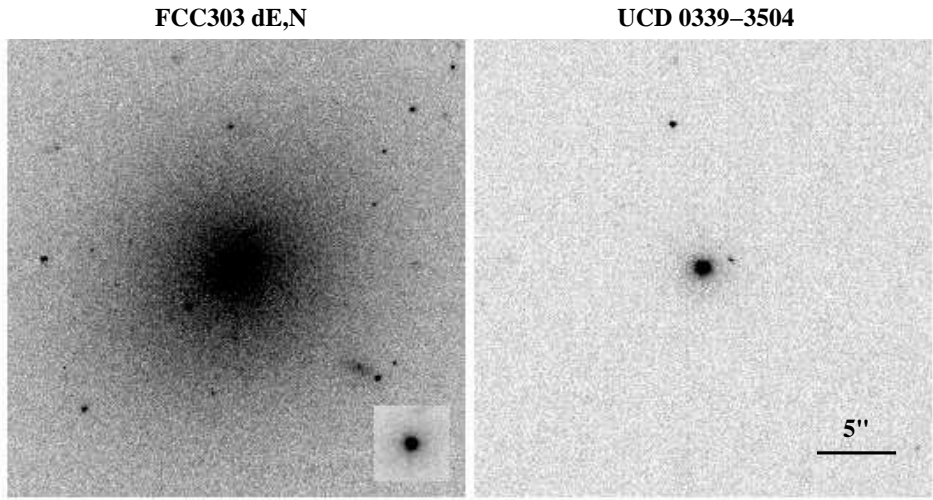}
\caption {STIS {\sc clearpass} images comparing a nucleated
dE to one of the UCDs.  Exposure times and display contrast are the
same; the fragile dE halo overfills the displayed region while the UCD
is bare.  The inset of the left panel shows the dE nucleus at lower
contrast to reveal its similarity to the UCD.  One arcsecond is about 100pc at the
distance of Fornax.}
\end{figure}

\vspace {-0.25in}
\section{Relation of Ultra Compact Dwarfs to Other Galaxies}
\vspace {-0.15in}

Though tiny and hard to find, the UCDs provide interesting insights
into the longer term evolution of galaxies in a dense environment.
Our favored interpretation of UCDs is that they are the durable
central nuggets of nucleated dwarf ellipticals which have had their
fragile, puffy halos tidally stripped or ``threshed'' by repeated
close and intense encounters with the largest galaxies in the cluster
(Bekki, et al.\ 2001a).  Both the stripped halos, accounting for
$\sim~98\%$ of a dE's total light (Binggeli \& Cameron 1991; Freeman
1993), and the surviving nuclei are dispersed into intracluster space
or added to the envelopes of the brighter galaxies, especially cDs
like NGC~1399.  Such phenomena are not unique to rich clusters;
tidally liberated streams of stars have been recognized around M31 and
the Milky Way (e.g., Ibata et al.\ 2001a,b), and can perhaps
account for the origin of M32 (Bekki et al.\ 2001b).

\begin{figure}[t!]
\includegraphics[width=4.5in,height=3.75in]{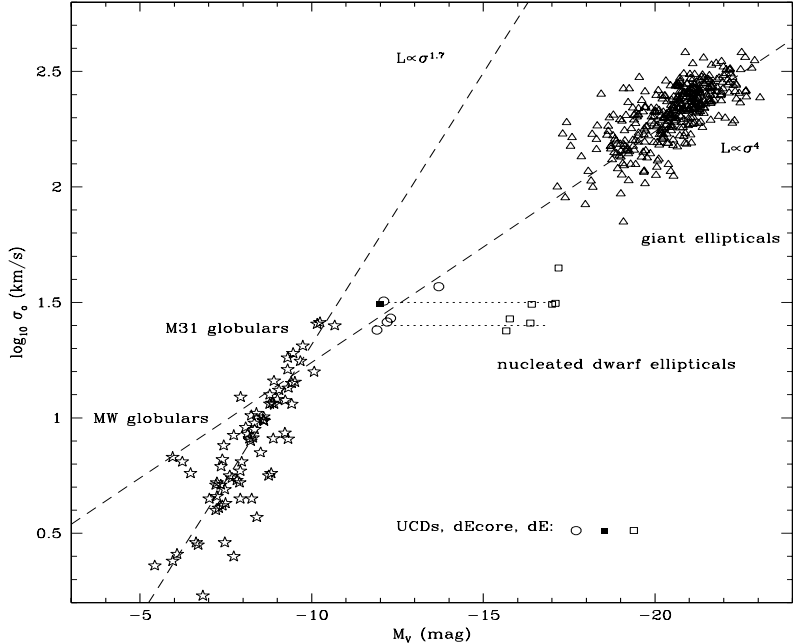}
\vspace{-0.5cm}
\caption{ Comparison of various early type objects, showing that UCDs
may be more closely related to dE,N galaxies than to bona fide
globular clusters.  Sources for velocity dispersions are: UCDs are
from our VLT echelle spectra; bright galaxies from Faber et al.\ 1987;
dE,N data are from Geha et al.\ (2001) (Virgo dEs) or our own spectra
for several Fornax dEs, including FCC303 (Figure~2) which was observed
with the VLT echelle; M31 globulars from Djorgovski et al.\ (1997). A
two component fit to the dE nucleus+halo has been used to estimate
their separate luminosities to place the dE nucleus separately in the
plot; it falls amongst the UCDs and is indicated by the tiny black
filled square.  }
\end{figure}

We have obtained HST STIS images of 5 UCDs
plus a comparison dE in Fornax.  These images do just resolve the UCD
cores and show that they are structurally a good match to the nucleus
of the comparison dE, after model subtraction of the dE halo
(Figure~2).  
To further investigate the possible links among UCDs, dEs, and globular
clusters, we have obtained echelle spectra of 5 of the UCDs at the
VLT using UVES and Keck with ESI.  When placed in
the M$_V$--$\sigma$ plane along with ellipticals, dEs, and
Galactic+M31 globular clusters (Figure~3), the UCDs appear to be
more closely related to brighter ellipticals than to globulars.  The
bright early types follow the well known $L \propto \sigma^4$
Faber-Jackson (1976) relation, while the globulars lie along a
different locus with $L \propto \sigma^{1.7}$ (Djorgovski et al.\
1997).  The UCDs, though lying closer to globulars in this plane, fall
along the extrapolated Faber-Jackson relation and are an order of
magnitude too bright for their velocity dispersions to join the
globular cluster relation.

\vspace {-0.25in}
\section{Role of UCDs in Galaxy Cluster Evolution}
\vspace {-0.15in}

If the threshing model of Bekki et al.\ (2001a) is correct and the UCDs
have lost their envelopes -- 98\% of their light -- while the nuclei
are little affected, then the UCDs would have begun life $\sim 4.25$
magnitudes brighter, in the regime of the nucleated dwarf ellipticals
(Figure~3).  The UCDs may hold the answer to the old puzzle of why cD
and some other cluster ellipticals have such high ``specific
frequencies'' of globular clusters, factors of several or more greater
than globular populations around spirals like the Milky Way.  During
the gravitational interactions which remove their halos, many UCDs may
be captured by the central elliptical where they can masquerade as
globulars, accounting for the brightest cD halo star clusters.
Threshing may also be able to explain the rest of the specific
frequency excess in cDs.  Figure~2 shows FCC303, a typical dE; in its
halo are $\sim 10$ globulars (seen as point sources at the distance of
Fornax).  The specific frequency of dEs is high (Miller et al.\ 1998),
comparable to that of a typical cD.  Recycled dE halo material
accumulated by a cD over a Hubble time provides a natural explanation
for the large globular cluster populations of giant ellipticals in
clusters.  Through such tidal interactions in rich environments
the evolution of bright galaxies and the cluster as a whole are
intimately linked to the population of dwarf galaxies.

\acknowledgements

\enlargethispage{2cm}

The authors acknowledge generous support from the National Science
Foundation (AST~9970884), and NASA through STScI.
Part of the work reported here was done at IGPP, under the
auspices of the U.S. Department of Energy by Lawrence Livermore
National Laboratory under contract No.~W-7405-Eng-48.

\end{article}

\vspace {-0.25in}
\begin{thebibliography}{}
\vspace {-0.1in}
\bibitem{} Bekki, K., Couch, W.J., \& Drinkwater, M.J. 2001a, ApJL, 552, 105
\bibitem{} Bekki, K., Couch, W.J., Drinkwater, M.J., \& Gregg, M.D. 2001b,
ApJL, 557, 39
\bibitem{} Binggeli B., Cameron L.M., 1991, A\&A, 252, 27
\bibitem{} Djorgovski, S. G., et al.\ 1997, ApJL, 474, L19
\bibitem{} Drinkwater, M.J., Phillipps, S., Jones, J.B., Gregg, M.D.,
Deady, J.H., Davies, J.I., Parker, Q.A., Sadler, E.M., \& Smith, R.M.,
2000a, A\&A, 355, 900
\bibitem{} Drinkwater, M.J., Jones, Gregg, M.D., \& Phillipps, S., 2000b
Pub. of the Astron. Soc. of Australia, 17, 227
\bibitem{} Faber, S.M. \& Jackson, R.E. 1976, ApJ, 204, 668
\bibitem{} Faber, S.M. et al.\ 1987, ApJS, 69, 763
\bibitem{} Ferguson, H.C. 1989, AJ, 98, 367
\bibitem{} Freeman, K.C. 1993, in ASP Conf.\ Ser.\ 48, The Globular
Cluster--Galaxy Connection, ed. G.H. Smith \& J.P. Brodie (San
Francisco: ASP), 608
\bibitem{} Geha, M., Guhathakurta, P., \& van der Marel, R. 2001, astro-ph/0107010
\bibitem{} Gregg, M.D. \& West, M.J. 1998, Nature, 396, 549 
\bibitem{} Ibata, R., Lewis, G. F., Irwin, M., Totten, E., \& Quinn,
2001a, ApJ, 551, 294
\bibitem{} Lopez-Cruz, O., et al.\ 1997 ApJL 475, L97
\bibitem{} Moore, B., Katz, N., Lake, G., Dressler, A., \& Oemler, A.,
Jr. 1996, Nature, 379, 613
\bibitem{} Miller, B. W., Lotz, J. M. Ferguson, H. C., Stiavelli, M.
\& Whitmore, B. C. 1998, ApJ, 508, 133
\bibitem{} Phillipps S., Drinkwater M.J., Gregg M.D., Jones, J.B., 2001,
ApJ, 560, 201
\end{thebibliography}
\end{document}